\documentclass[10pt]{article}
\usepackage{graphicx}
\usepackage{lineno}

\def\Title#1{\begin{center} {\Large #1 } \end{center}}
\def\Author#1{\begin{center}{ \sc #1} \end{center}}
\def\Address#1{\begin{center}{ \it #1} \end{center}}

\newcommand\pubblock{\rightline{\begin{tabular}{l} Proceedings of the Fifth Annual LHCP\\ \pubnumber\\
         \pubdate  \end{tabular}}}

\newenvironment{Abstract}{\begin{quotation} \begin{center} 
             \large ABSTRACT \end{center}\bigskip 
      \begin{center}\begin{large}}{\end{large}\end{center} \end{quotation}}

\newenvironment{Presented}{\begin{quotation} \begin{center} 
             PRESENTED AT\end{center}\bigskip 
      \begin{center}\begin{large}}{\end{large}\end{center} \end{quotation}}

\def\beq{\begin{equation}}
\def\eeq#1{\label{#1}\end{equation}}
\def\eeqn{\end{equation}}

\def\beqa{\begin{eqnarray}}
\def\eeqa#1{\label{#1}\end{eqnarray}}
\def\eeqan{\end{eqnarray}}

\let\bar=\overbar

\def\Dslash{\not{\hbox{\kern-4pt $D$}}}
\def\dslash{\not{\hbox{\kern-2pt $\del$}}}

\def\msb{{\bar{\ssstyle M \kern -1pt S}}}

\usepackage{color}
\usepackage{subfigure}
\usepackage{cite}
\usepackage{wrapfig}
\newcommand\pubnumber{ ATL-PHYS-PROC-2017-096 }
\newcommand\pubdate{\today}

\def\affiliation{
On behalf of the ATLAS Collaboration, \\
International Center for Elementary Particle Physics and Department of Physics, The University of Tokyo, Tokyo, Japan}

\begin{document}
%\linenumbers
% large size for the first page
\large
\begin{titlepage}
\pubblock

%% Change the title, name, abstract
%% Title 
\vfill
\Title{  SUSY strong production in leptonic final state with ATLAS  }
\vfill

%  if you need to add the support use this, fill the \support definition above. 
%   \Author{ FIRSTNAME LASTNAME \support }
\Author{ Tomoyuki Saito  }
\Address{\affiliation}
\vfill
\begin{Abstract}
Supersymmetry is one of the most motivated scenarios for physics beyond the Standard Model.
This article summarizes recent ATLAS results on searches for supersymmetry in proton-proton collisions at a centre-of-mass energy of 13~TeV at LHC, which target supersymmetric particles produced by strong interaction in events with leptonic final states.
No significant excess above the Standard Model expectation is observed and exclusion limits have been set on squark and gluino masses in various scenarios.

\end{Abstract}
\vfill

% DO NOT CHANGE 
\begin{Presented}
The Fifth Annual Conference\\
 on Large Hadron Collider Physics \\
Shanghai Jiao Tong University, Shanghai, China\\ 
May 15-20, 2017
\end{Presented}
\vfill
\end{titlepage}
\def\thefootnote{\fnsymbol{footnote}}
\setcounter{footnote}{0}
%

% normal size for the rest
\normalsize 

%% Your paper should be entered below. 

\section{Introduction}
The Standard Model~(SM) of particle physics is extremely successful in describing the phenomena of elementary particles and their interactions.
Nevertheless it is believed to be a low energy realisation of a more general theory, because of the problems, inside the SM, related on the stabilisation of the Higgs mass against radiative corrections from Planck scale physics.
The shortcoming could be explained by new physics appearing at the TeV scale.

Supersymmetry~(SUSY)\cite{SUSY1,SUSY2,SUSY3,SUSY4,SUSY5,SUSY6} is a generalization of space-time symmetries that predicts new bosonic partners for the fermions and new fermionic partners for the bosons of the SM whose spins differ by half unit.
The partners of the fermionic particles~(quarks and leptons) are the scalar squarks~($\tilde{q}$) and sleptons~($\tilde{l}$).
The supersymmetric partner of the gluon is the fermionic gluino~($\tilde{g}$).
The supersymmetric partner of the higgs boson~(higgsino) and the electroweak gauge bosons~(bino and winos) mix to form charged mass eigenstates~(chargino, $\tilde{\chi}^{\pm}_{1,2}$) and neutral mass eigenstates~(neutralino, $\tilde{\chi}^0_{1,2,3,4}$).
The lightest neutralino $\tilde{\chi}^0_{1}$ could be the lightest SUSY particle~(LSP) and a good dark-matter candidate.
The SUSY also provides a solution to the Higgs mass problem.

Large cross-sections predicted for the strong production of supersymmetric particles make the production of gluinos and squarks a primary target in searches for SUSY at LHC.
%% Final state
In R-parity conserved SUSY\cite{SUSY_RPC}, the gluinos and squraks can be pair-produced and decay either directly or via intermediate states to the LSP, emitting jets and leptons.
The kinematics of the jets and leptons strongly depends on mass differences between the SUSY particles in a given decay.
Aiming to cover all possible scenarios, our search of SUSY is based on topology of the final state, not optimized for a specific scenario.
%% This document organization
In this article, we discuss recent searches of SUSY strong production with leptonic final states at ATLAS\cite{ATLAS}.
The searches use the ATLAS data collected in proton-proton collisions in 2015 and 2016 at a centre-of-mass energy of 13~TeV.
Firstly, the searches for R-parity conserved decays are described in Section \ref{sec: searchRPC}, in which  SUSY particles are produced in pairs, and decay into multiple-jets, ($b$-tagged jets,) LSPs and one or more leptons.
Secondly, the search for R-parity violated decays is described in Section \ref{sec: searchRPV}, in which the SUSY particles are produced in pairs, and decay into multiple-jets, ($b$-tagged jets,) and one lepton.

\section{Search Strategy}
\label{sec:searchstrategy}
\begin{wrapfigure}[20]{r}[0pt]{0.40\textwidth}
\vspace{-0.5cm}
\centering
\includegraphics[width=0.40\textwidth]{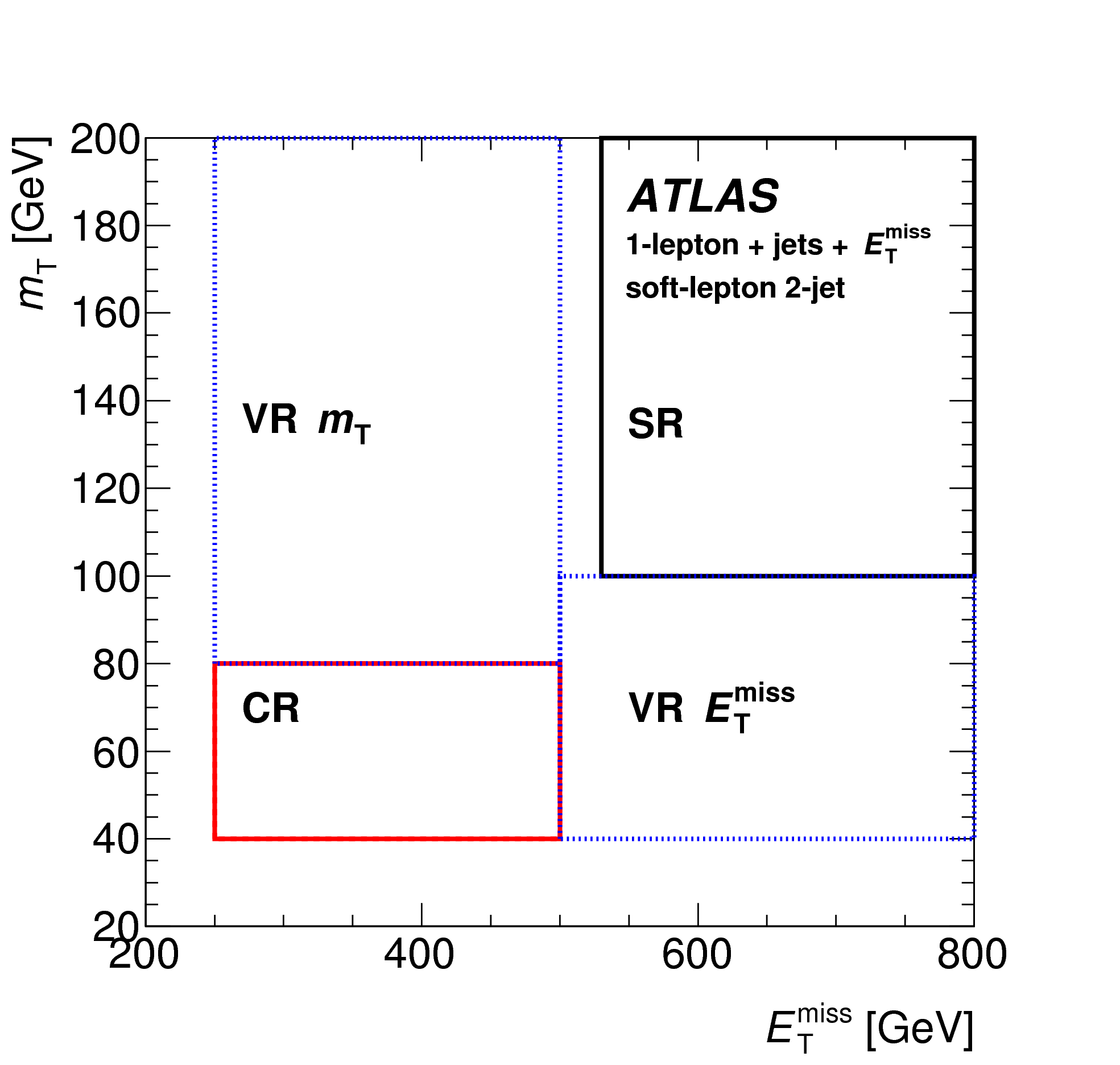}
\vspace{-0.5cm}
\caption{ An example of the definitions of signal region~(SR), control region~(CR) and validation region~(VR). }
\label{fig:RegionDesign}
\end{wrapfigure}
Typical SUSY searches are performed in corners of phase space with a large number of jets, large missing transverse momentum~( $E_{\rm T}^{{\rm miss}}$) and large $m_{{\rm eff}}$ defined as a scalar sum of the $p_{\rm T}$ of the objects and $E_{\rm T}^{{\rm miss}}$ in the final state.
Background estimation is challenging in the search of SUSY.
The dominant backgrounds from the SM process are estimated via partially data-driven techniques with dedicated control regions~(CR).
The CR is enriched in the background process of interest and designed to be kinematically close but orthogonal to the regions with signal regions~(SR).
The Monte Carlo~(MC) simulations are normalized to data in the CR and extrapolated to the SR with shape information obtained from MC.
The MC modeling in the extrapolation are tested in validation regions~(VR), which is designed to sit kinematically between the CR and SR and typically validate the extrapolation of a single variable.
Figure \ref{fig:RegionDesign} shows an example for the definition of SR, CR and VRs.
The VR is defined such that background events there show similar kinematics properties as in the SR.
Minor backgrounds are typically taken directly from MC.

In each SR, the event yield is compared with the expectation from the SM processes, which is estimated using a combination of simulation and observed data in CRs.

\clearpage

\section{Searches for SUSY strong production with leptonic and R-parity conserved decay}
\label{sec: searchRPC}
%%%
%%% One-Lepton
%%%
\subsection{One-lepton final state}
\label{subsec:RPC_onelep}
\begin{wrapfigure}[21]{r}[0pt]{0.30\textwidth}
\centering
\includegraphics[width=0.26\textwidth, bb=0 0 130 109]{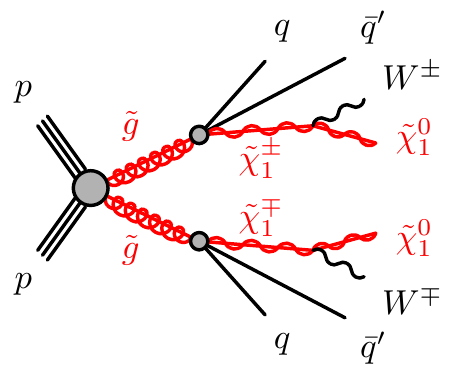}
\includegraphics[width=0.26\textwidth, bb=0 0 130 109]{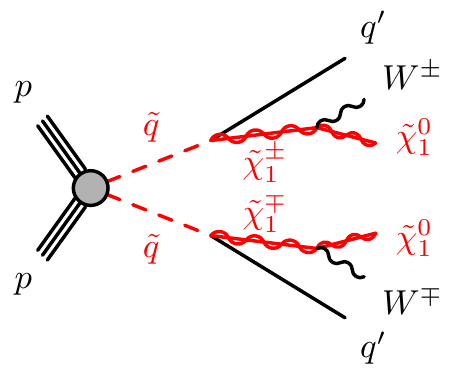}
\vspace{-0.2cm}
\caption{Diagrams of the decays targeted by one-lepton search }
\label{fig:onestep_diagram}
\end{wrapfigure}

%%% Final state and motivation
Final states in this search consist of a lepton~(electron and muon), multiple jets and a large $E_{\rm T}^{{\rm miss}}$ from the undetectable LSP~($\tilde{\chi}_{1}^{0}$).
The benchmark decays are illustrated at Figure \ref{fig:onestep_diagram}.
The search uses the ATLAS data in 2015 and 2016 corresponding to an integrated luminosity of 14.8~fb$^{-1}$\cite{RPC_onelep}.

%%% SR
The number of jets and their kinematics depends on the mass difference between the SUSY particles in the decay.
Ten sets of signal regions are defined in 2-, 4-, 5- and 6-jet selections in order to cover all possible phase spaces.
The 2-jet SR provides a sensitivity to scenarios characterised by small mass differences between $m_{\tilde{g}}$, $m_{\tilde{\chi}_{1}^{\pm}}$ and $m_{\tilde{\chi}_{1}^{0}}$, where most of the decay products tend to have small $p_{\rm T}$. 
The 6-jet SR targets models with large mass differences, where the decay products have relatively high $p_{\rm T}$. 
The 4- and 5-jet SRs provides a sensitivity to scenarios with intermediate mass differences between the 2- and 6-jet SRs.
A suppression of background, in addition to typical variables like $m_{{\rm eff}}$, is based on a difference on the event shape between the SUSY signal and SM background; the aplanarity \cite{aplanarity}.
The signal events tend to be more spherical than background events.

%%% Background
Dominant SM backgrounds in the SRs are $t\overline{t}$ and $W+$jets.
CRs are defined as low $m_{\rm T}$ and aplanarity or $E_{\rm T}^{{\rm miss}}$ region to normalize the MC to data, and  extrapolation modelings to the SRs are tested at VRs.

\begin{figure}[htb]
\centering
\hspace{-1.0cm}
%\subfigure{\includegraphics[width=0.23\textwidth, bb=0 0 1418 1360]{mT_GG6Jbulk.png}
\subfigure{\includegraphics[width=0.30\textwidth]{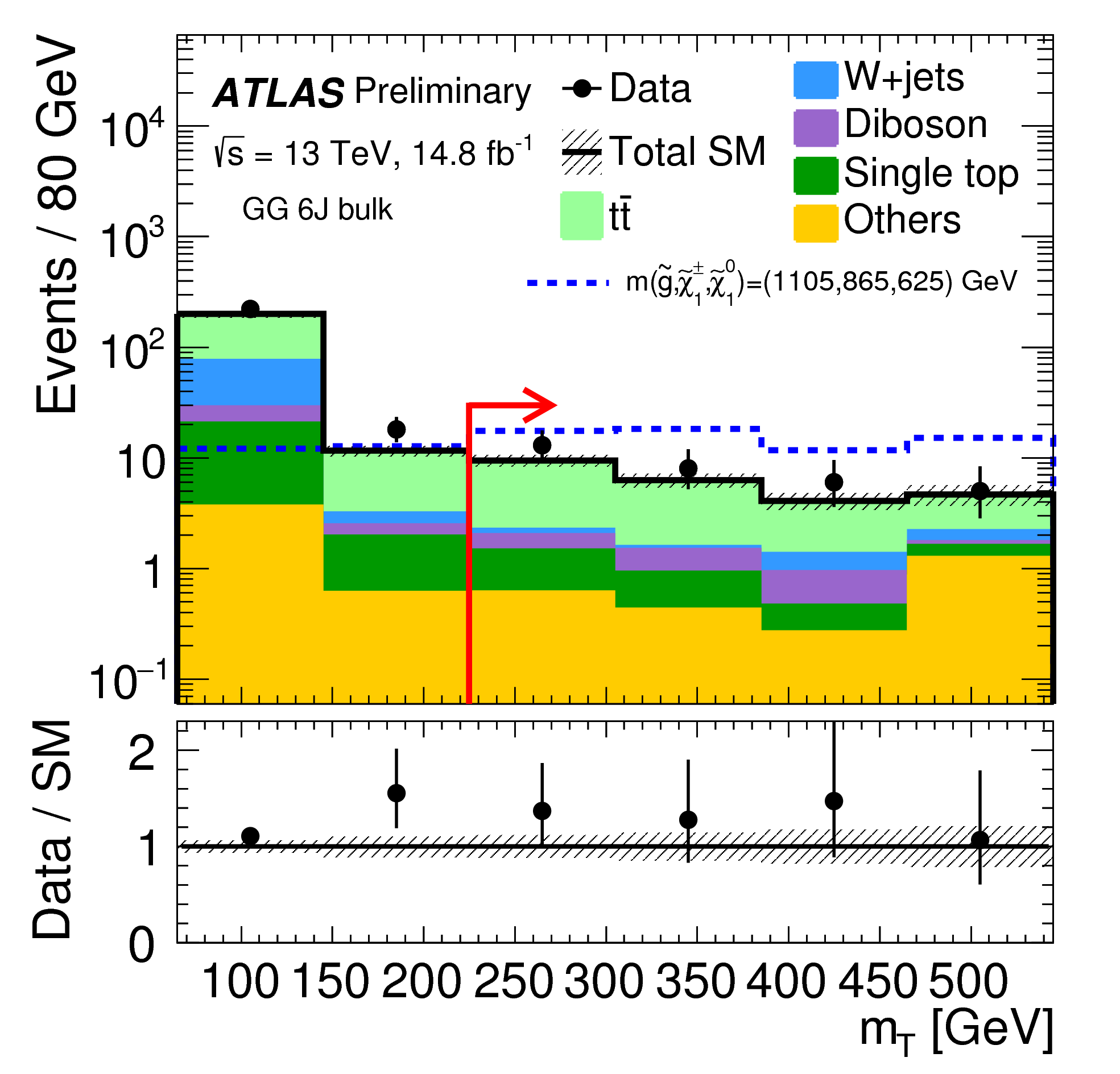}\label{fig:RPC_onestep_SRplot1}}
%\hspace{0.5cm}
\subfigure{\includegraphics[width=0.30\textwidth]{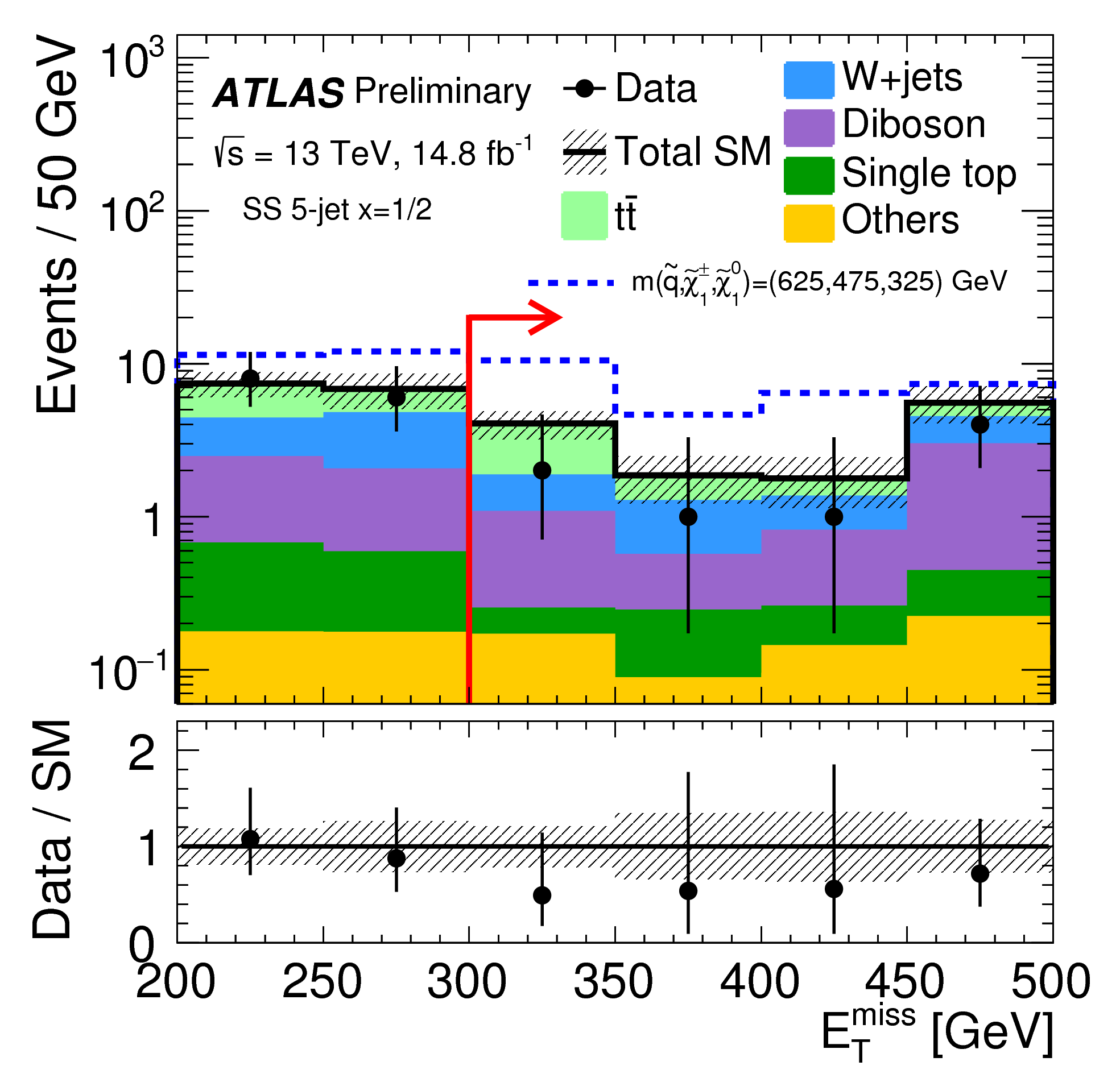}
%\subfigure{\includegraphics[width=0.23\textwidth, bb=0 0 1418 1360]{ETmiss_SS5J.png}
\label{fig:RPC_onestep_SRplot2}}
%\hspace{0.5cm}
\subfigure{\includegraphics[width=0.41\textwidth]{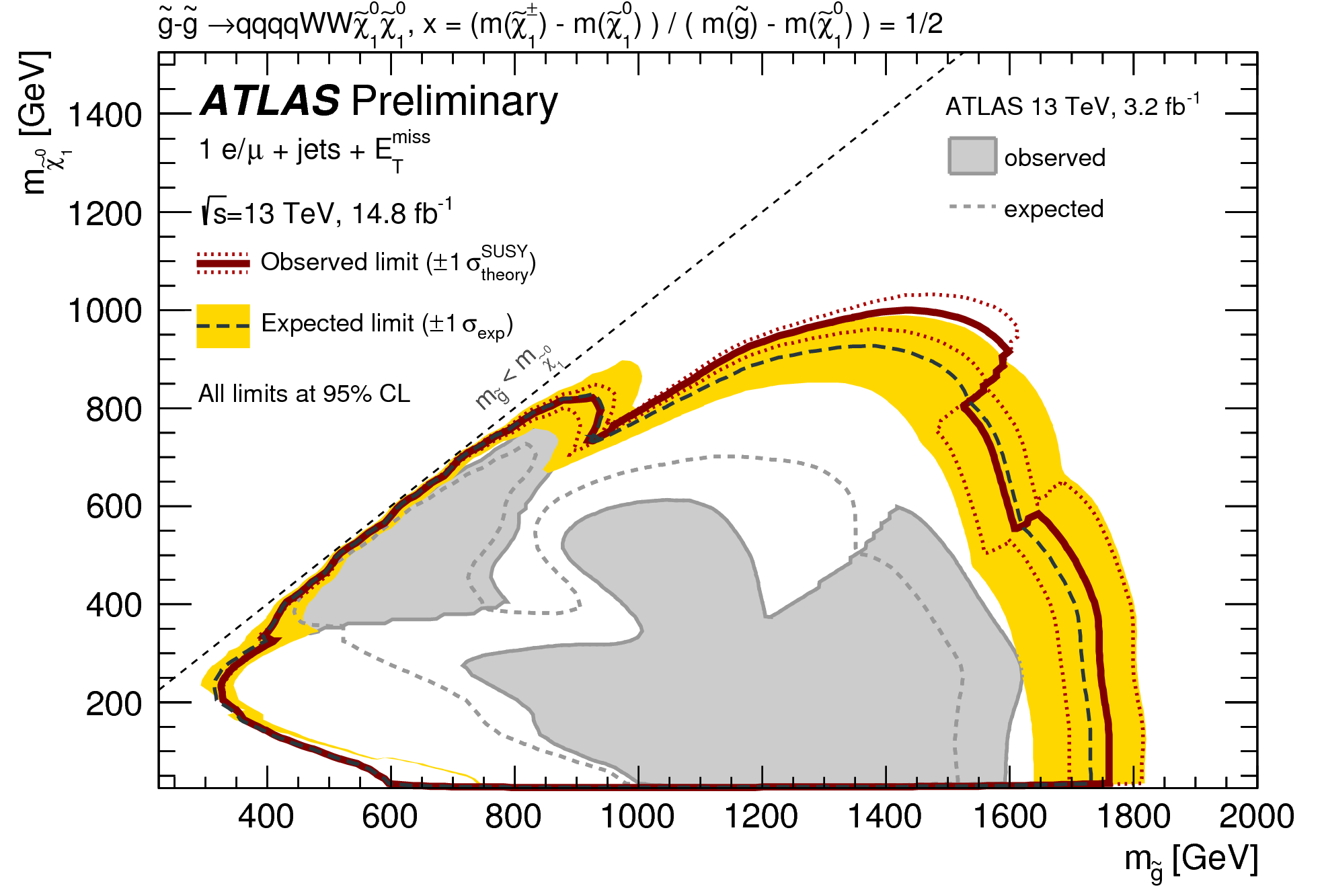}
%\subfigure{\includegraphics[width=0.33\textwidth, bb=0 0 1418 960]{onelepRPC_onestepGG_limit.png}
\label{fig:RPC_onestep_limit}}
\caption{ (Left): Transverse mass~($m_{\rm T}$) distribution between the lepton and $E_{\rm T}^{{\rm miss}}$, for the gluino 6-jets SR. (Middle): $E_{\rm T}^{{\rm miss}}$ distribution for the squark 5-jets SR. (Right): The combined 95\% CL exclusion limits. The signal region with the best expected sensitivity is used for each model point. \cite{RPC_onelep}}
\label{fig:onelep_result}
\end{figure}

%%% Result
Comparisons between data and fitted background estimates are shown in Figure \ref{fig:onelep_result}~(left and middle) for 6-jets and 5-jets SRs, respectively.
The background estimates agree well with the data in all regions, and no significant excess is observed.
Figure \ref{fig:onelep_result}~(right) shows the combined 95\% CL exclusion limits in the gluino decay, where the signal region with the best expected sensitivity is used for each model point.

\vspace{0.5cm}

\clearpage

%%%
%%% Multi-b One-Lepton
%%%
\subsection{One-lepton + Multiple $b$-tagged jets final state}

\begin{figure}[htb]
\centering
%\vspace{0.3cm}
\hspace{-0.5cm}
\subfigure{\includegraphics[width=0.27\textwidth]{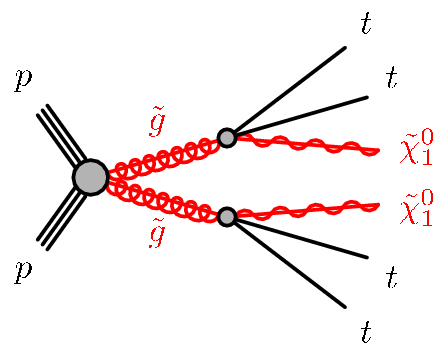}}
%\subfigure{\includegraphics[width=0.20\textwidth, bb=0 0 320 255]{gtt_diagram.png}}
\hspace{0.5cm}
%\hspace{1.5cm}
\subfigure{\includegraphics[width=0.32\textwidth]{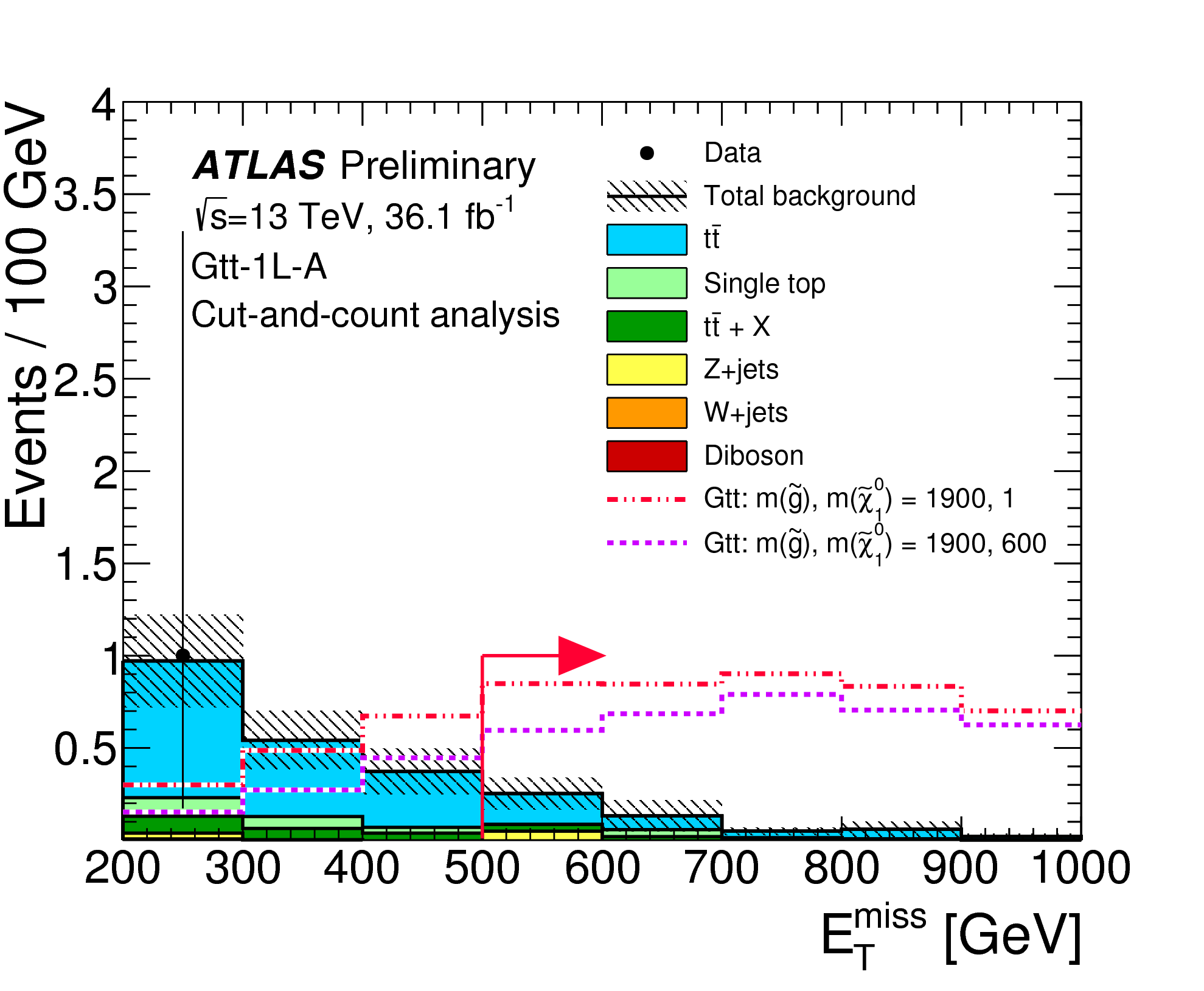}}
%\subfigure{\includegraphics[width=0.24\textwidth, bb=0 0 1425 1173]{Gtt_SRA_ETmiss.png}}
%\hspace{1.5cm}
%\subfigure{\includegraphics[width=0.28\textwidth, bb=0 0 1425 968]{Gtt_limit.png}}
\subfigure{\includegraphics[width=0.35\textwidth]{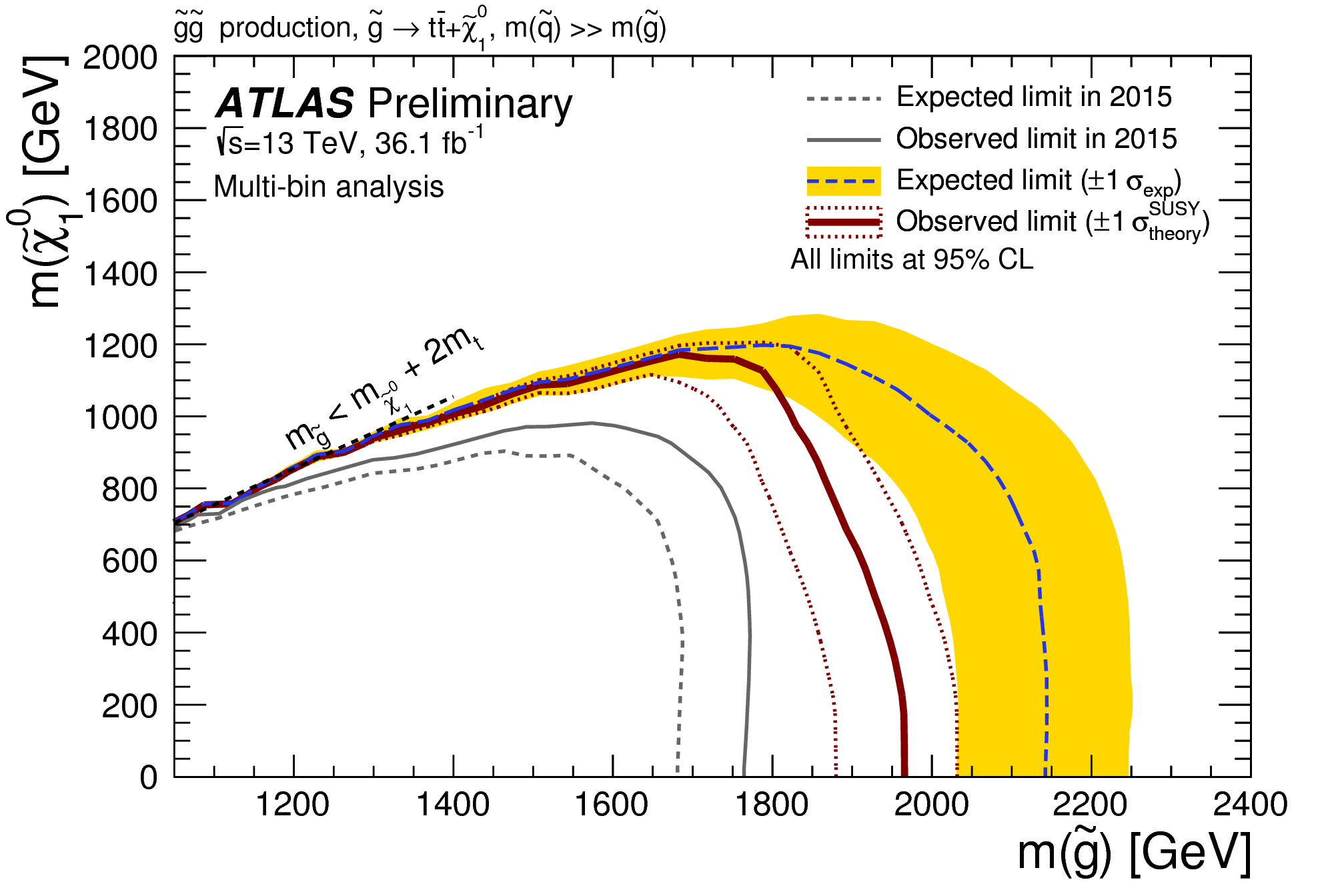}}
\vspace{-0.5cm}
\caption{ (Left): The decay topology for the final state with a lepton and multiple $b-$tagged jets.  (Middle): Distribution of $E_{\rm T}^{{\rm miss}}$ for the SR-A. (Right): The 95\% CL observed and expected exclusion limits is shown in the LSP and gluino mass plane. \cite{multib}}
\label{fig:multib_result}
\end{figure}

%%% Final state and motivation
The SUSY can suppress the scale hierarchy reducing unnatural tuning in the Higgs sector, provided that the scalar top partner has a mass not too far above the weak scale.
Since the gluinos are expected to be pair-produced with a large cross section at LHC the search for gluino production with decays via off-shell stop quarks shown in Figure \ref{fig:multib_result}~(left) is highly motivated.
This section presents the search for gluino pair production decaying via off-shell stop quark in events with multiple $b-$jets, high $E_{\rm T}^{{\rm miss}}$, and additional light quark jets and a lepton~(electron and muon) with 36.1~fb$^{-1}$ dataset \cite{multib}.

SR-A~($N^{{\rm jet}}\geq5$) is optimized for signals with a large mass difference between the gluino and the neutralino, possibly leading to highly boosted objects in the final state.
The decay products of a hadronically-decaying boosted top quarks can be reconstructed in a single large-radius reclustered jet.
SR-C~($N^{{\rm jet}}\geq7$) focuses on signals for which the gluino decay products are softer due to the small mass difference, and SR-B~($N^{{\rm jet}}\geq6$) targets intermediate regions of the mass difference.

%%% Result
No significant excess is found above the predicted background in all SRs.
The 95\% CL observed and expected exclusion limits is shown in the LSP and gluino mass plane in Figure \ref{fig:multib_result} (right).

%\vspace{0.2cm}
\vspace{-0.5cm}

%\clearpage
\vspace{-0.5cm}
%%%
%%% Two-Leptons/Three-Leptons
%%%
\subsection{Two-leptons~(same-sign) and Three-leptons final states}
%%% Final state and motivation

\begin{figure}[htb]
\centering
\subfigure{\includegraphics[width=0.24\textwidth]{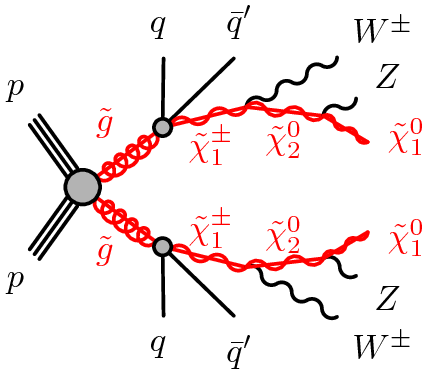}}
%\subfigure{\includegraphics[width=0.18\textwidth, bb=0 0 307 269]{2L3L_2step_diagram.png}}
\hspace{2.0cm}
\subfigure{\includegraphics[width=0.28\textwidth]{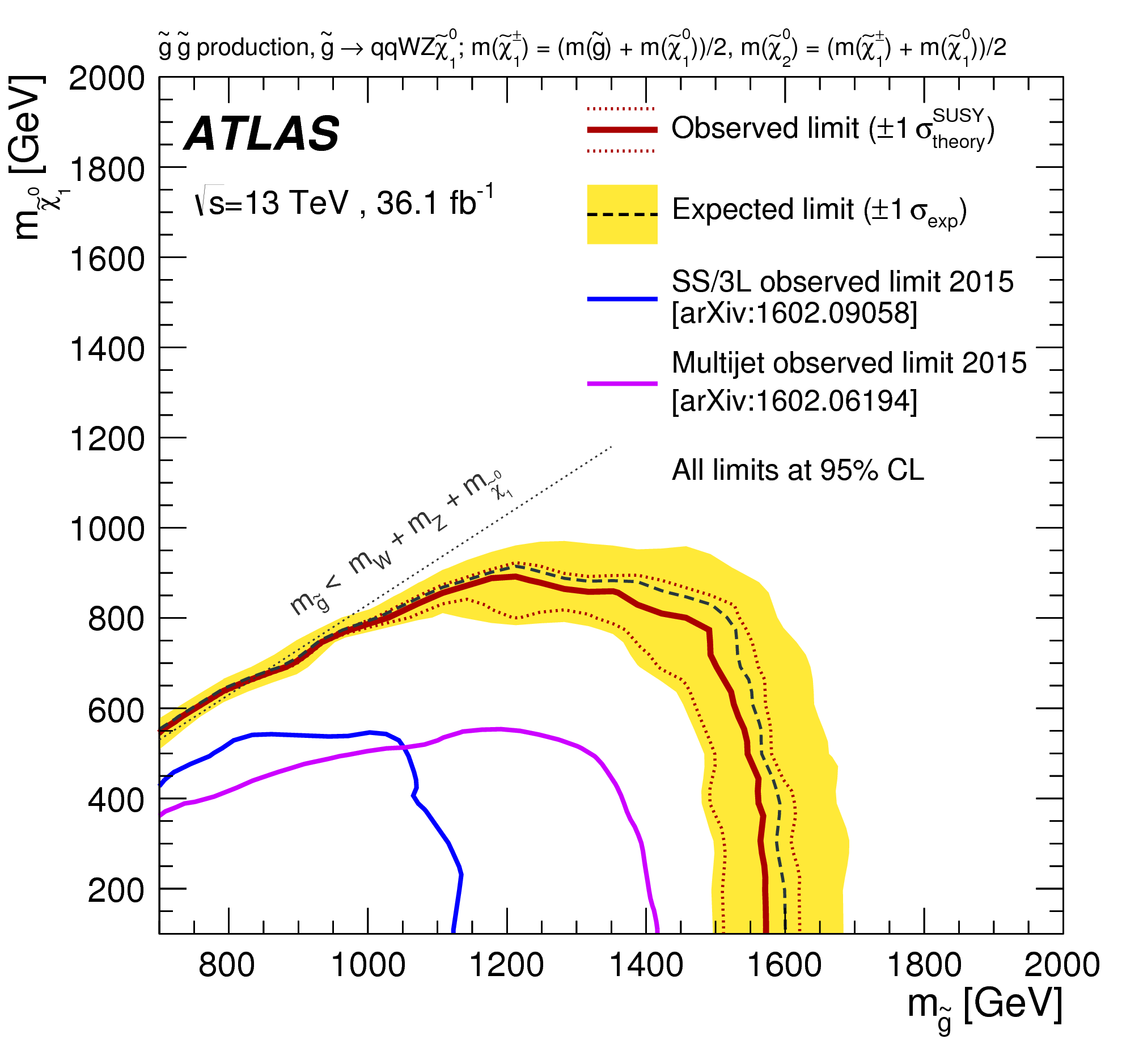}}
%\subfigure{\includegraphics[width=0.21\textwidth, bb=0 0 1425 1351]{2L3L_2step_limit-2.png}}
\vspace{-0.3cm}
\caption{ (Left): The decay diagram for the final state with the SS and 3S. (Right): The observed and expected exclusion limits on the masses of gluino and LSP\cite{2L3L}. }
\label{fig:SS3L}
\end{figure}

This section presents a search for SUSY in final states with two leptons (electrons or muons) of the same electric charge~(referred to as same-sign~(SS)) or three leptons~(3S), jets and $E_{\rm T}^{{\rm miss}}$ with 36.1~fb$^{-1}$ dataset\cite{2L3L}.
This search targets final states with multiple leptons from long decay chain, like Figure \ref{fig:SS3L}~(left).
SM processes leading to such final states have very small cross-sections.
Compared to other searches, the analyses based on same-sign or three leptons signatures allow a use of looser kenematic requirements, preserving a sensitivity to scenarios with small mass differences between gluinos/squarks and the LSP.

%%% Result
No significant excess is observed in all SRs.
Figure \ref{fig:SS3L}~(right) shows the exclusion limit on $\tilde{g}$ and $\tilde{\chi}_{1}^{0}$ masses.

%\vspace{0.8cm}
\vspace{-0.3cm}

\clearpage

%%%
%%% RPV One-Lepton
%%%
\section{Searches for SUSY strong production with leptonic and R-parity violated decay}
\label{sec: searchRPV}
\subsection{One-lepton final state}
\begin{wrapfigure}[32]{r}[0pt]{0.28\textwidth}
\vspace{-0.3cm}
\includegraphics[width=0.23\textwidth]{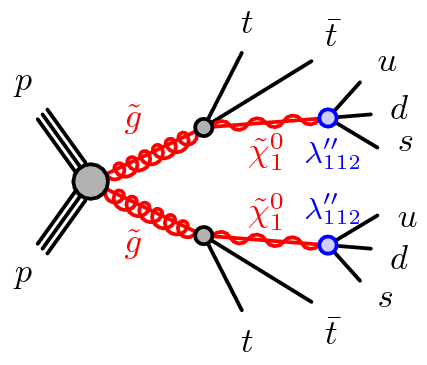}
\vspace{-0.3cm}
\caption{ The benchmark signal decay. }
\label{fig:RPV_onelep_diagram}
\vspace{0.2cm}
%\vspace{1.7cm}
\includegraphics[width=0.28\textwidth]{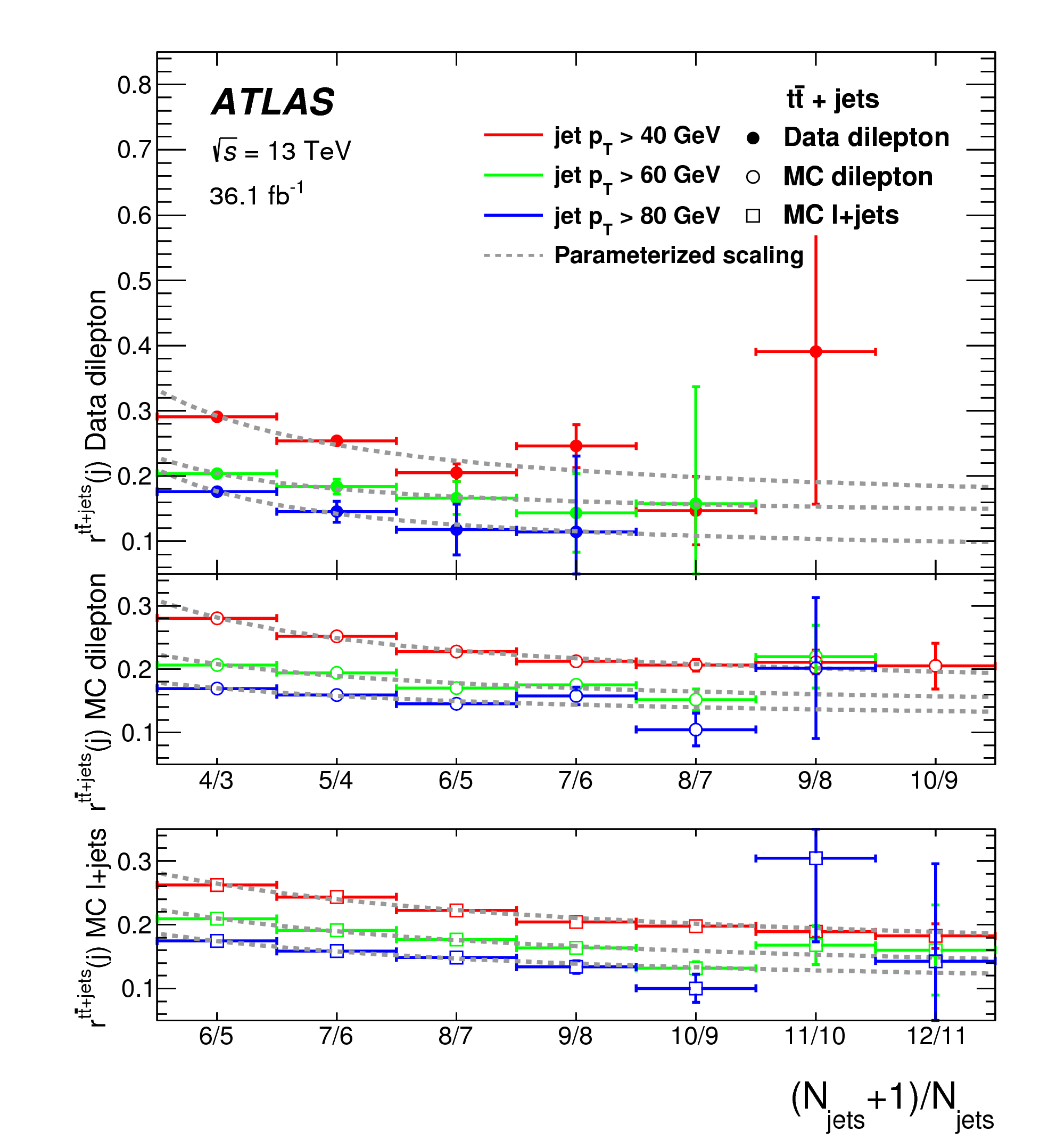}
\vspace{-0.8cm}
\caption{ The ratio of the number of events with ($j+$1) jets to that of $j$ jets\cite{RPV_onelep}. }
\label{fig:RPV_onelep_jetmodeling}
\end{wrapfigure}

%%% Final state and motivation
This search focuses on a final state with a lepton and multiple jets\cite{RPV_onelep}.
Unlike the SUSY searches of similar final states, no requirement on the $E_{\rm T}^{{\rm miss}}$ is applied. 
R-parity violating SUSY signatures (Figure \ref{fig:RPV_onelep_diagram}) are used for bench mark models in this search. 

%%% SR
The SRs are defined by requiring exactly zero or at least three $b-$tagged jets for a given minimum number of jets~(from 8- to 12-jets) with some jet $p_{\rm T}$ thresholds~(40, 60 and 80~GeV).

%%% Background
In this search, the modeling of the background at high jet multiplicity suffers from a large uncertainty.
A data-driven method is used to estimate the number of events in a given jet and $b$-tagged jet  multiplicity bin.
A concept of the background estimation is based on an extraction of the initial template of the $b$-tagged jet multiplicity distribution in events with five jets and a parameterization of the evolution of this template to higher jet multiplicities.
The extrapolation of the $b-$tag multiplicity distribution to higher jet multiplicities is based on an assumption that a difference between the $b-$tag multiplicity distribution in events with $j$ and $j+1$ jets arises mainly from the production of additional jets, and can be described by a fixed probability that the additional jet is $b-$tagged.
Figure \ref{fig:RPV_onelep_jetmodeling} presents a comparison of the scaling behaviour in data and MC simulation compared to a fit of the parametrization used, and shows that the assumed function describes the data and MC simulation well for the jet-multiplicity range relevant to this search.
Main systematic uncertainties are related to this estimation.
These are estimated by studying the closure of the method in different MC samples, including alternative MC generators and varying the event selection.
The uncertainties assigned vary from 3\% to 60\%.

\begin{figure}[htb]
\centering
%\hspace{-3.0cm}
\hspace{-0.5cm}
\subfigure{\includegraphics[width=0.32\textwidth]{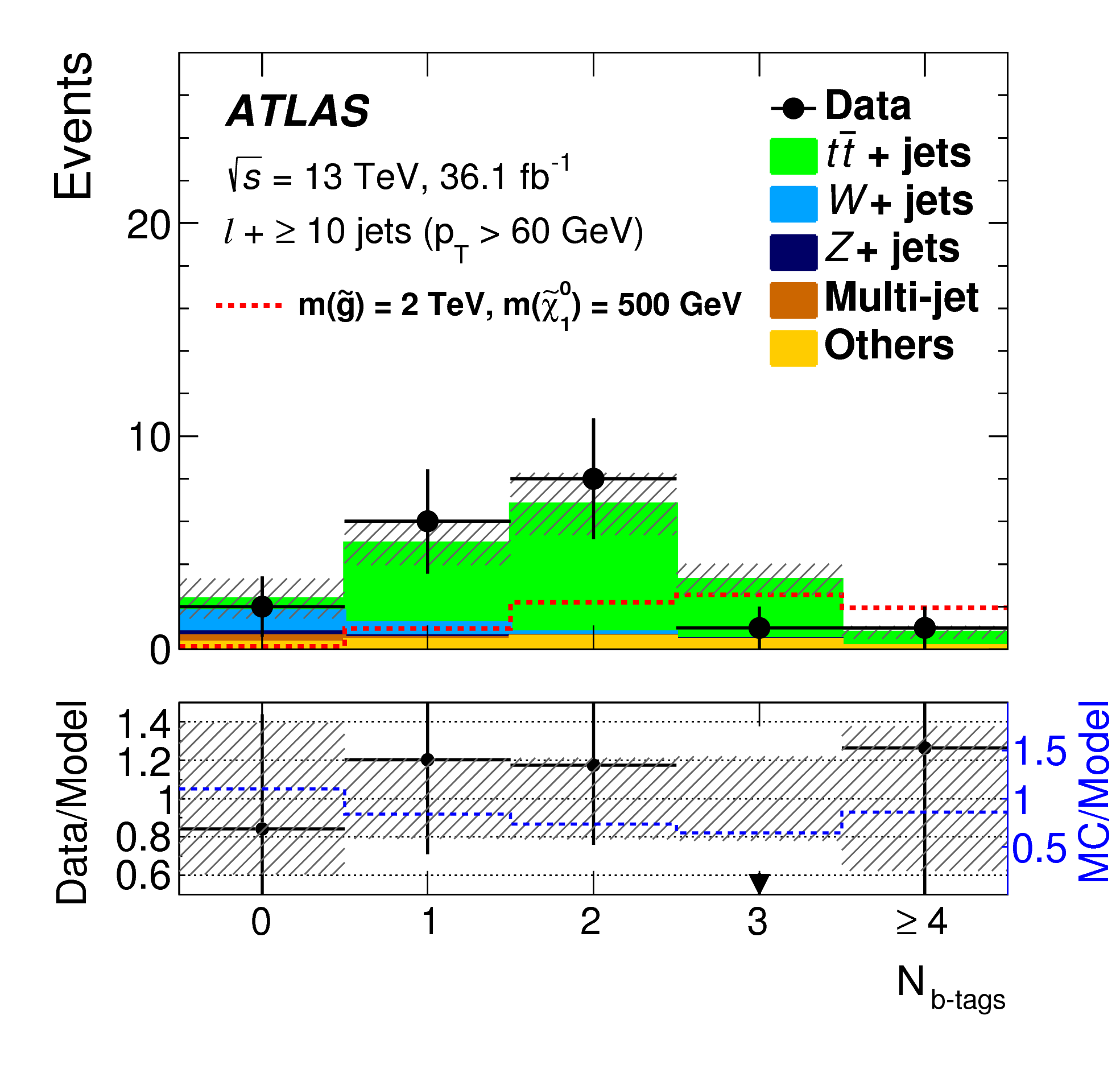}
%\subfigure{\includegraphics[width=0.27\textwidth, bb=0 0 1418 1348]{RPV_onelep_SR10J_nbjets.png}
\label{fig:RPV_onelep_SRplot1}}
%\hspace{1.5cm}
\subfigure{\includegraphics[width=0.32\textwidth]{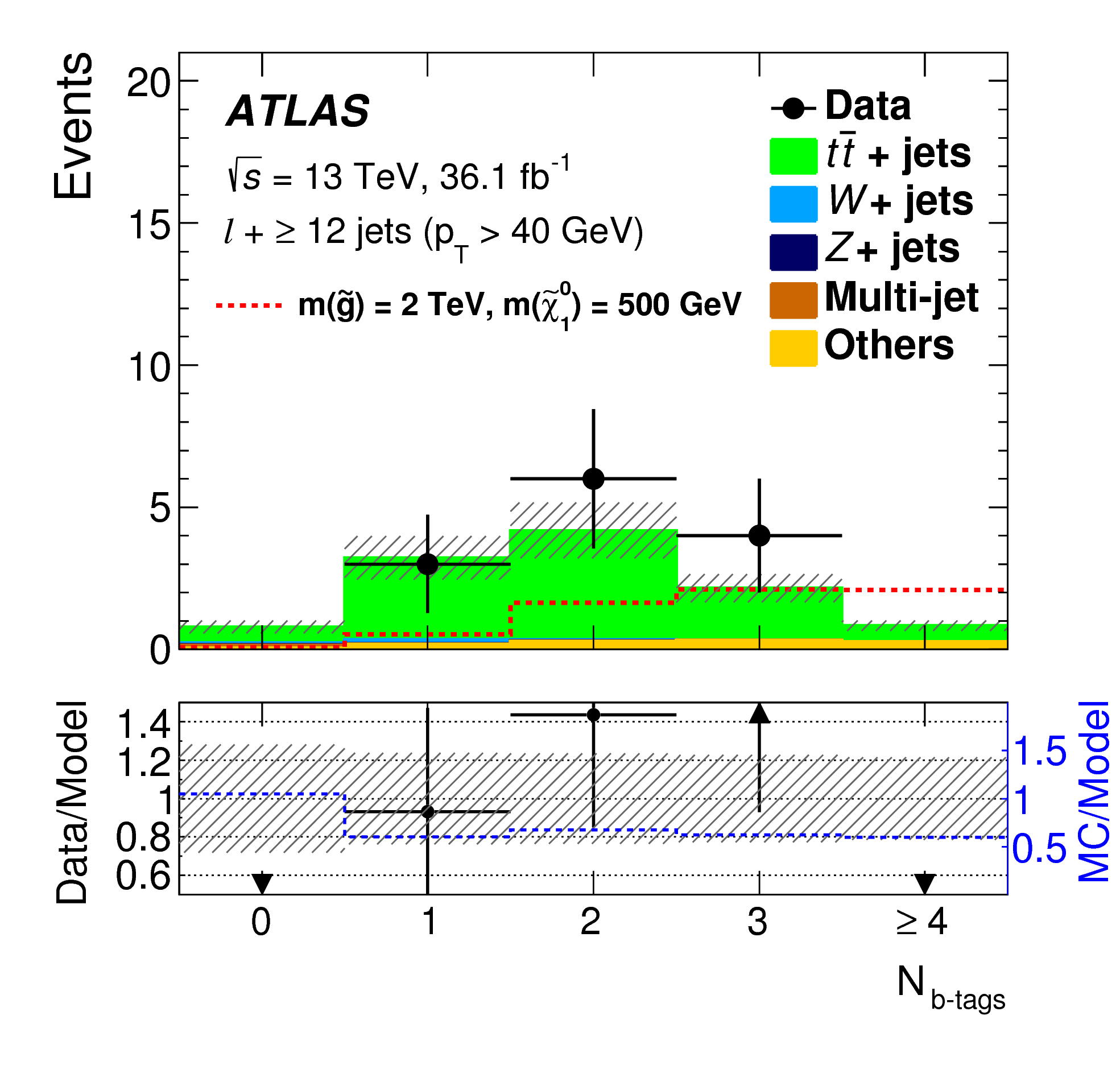}
%\subfigure{\includegraphics[width=0.27\textwidth, bb=0 0 1418 1348]{RPV_onelep_SR12J_nbjets.png}
\label{fig:RPV_onelep_SRplot2}}
%\hspace{1.5cm}
\subfigure{\includegraphics[width=0.32\textwidth]{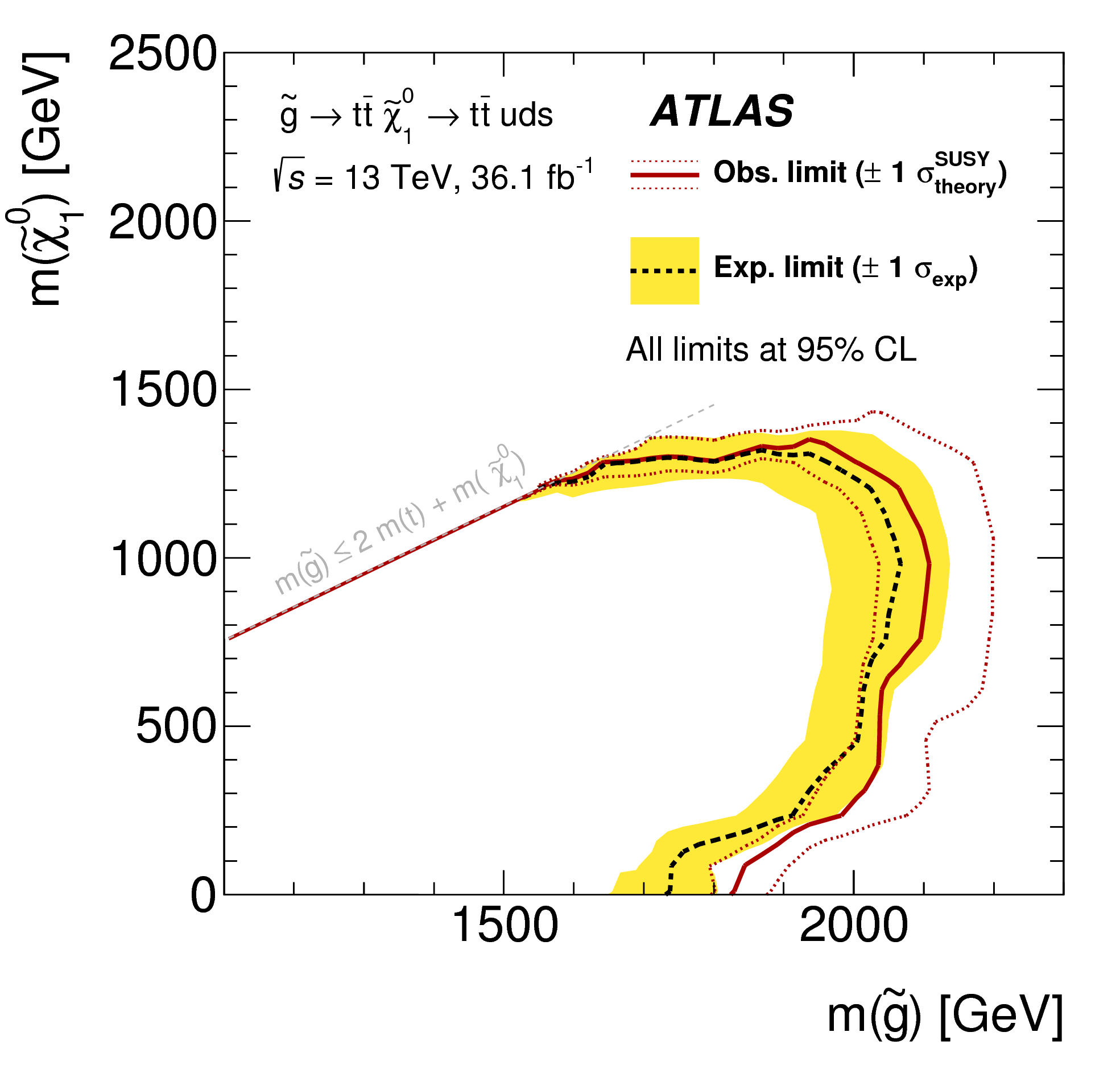}
%\subfigure{\includegraphics[width=0.27\textwidth, bb=0 0 1418 1552]{RPV_onelep_limit1.png}
\label{fig:RPV_onelep_limit}}
\vspace{-0.3cm}
\caption{ Results of the RPV one-lepton search. The expected background and observed data in the different jet and $b-$tagged jet  multiplicity bins for the 60~GeV (Left) and 40~GeV (Middle) jet $p_{\rm T}$ thresholds. (Right) : Observed and expected exclusion contours on the $\tilde{g}$ and $\tilde{\chi}^0_{1}$ masses in the context of the RPV SUSY scenario. \cite{RPV_onelep}}
\label{fig:RPV_onelep_SRplot}
\end{figure}

%%% Result
Figure \ref{fig:RPV_onelep_SRplot}~(left and middle) show the observed numbers of data events compared to the fitted background model.
No significant excess over the SM expectation is observed.
Figure \ref{fig:RPV_onelep_SRplot}~(right) shows the observed and expected exclusion limits in the plane of the gluino mass and neutralino mass.

\vspace{0.5cm}

\clearpage

\section{Conclusion}
\begin{figure}[htb]
\centering
\includegraphics[width=0.40\textwidth]{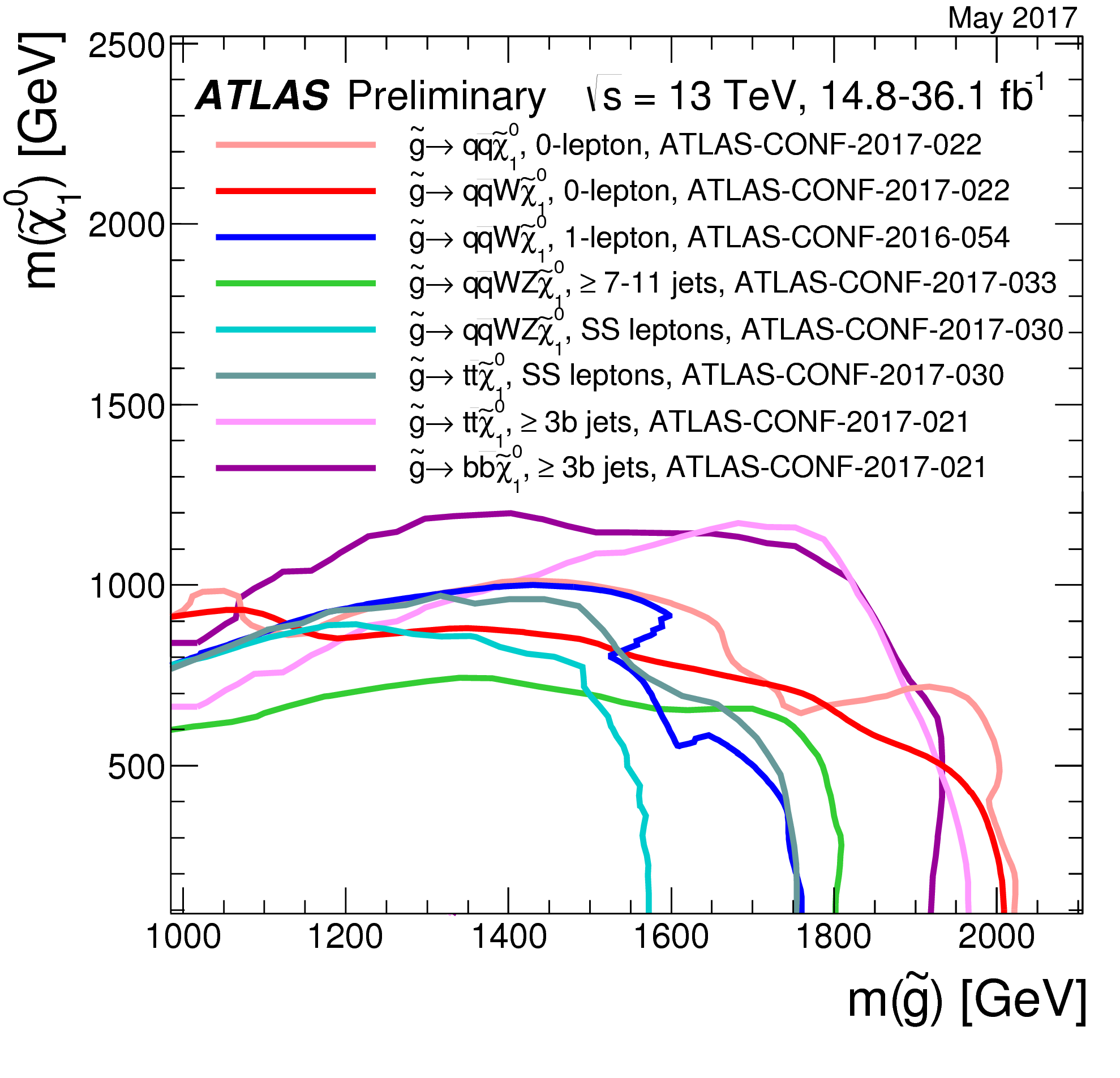}
%\includegraphics[width=0.30\textwidth, bb=0 0 1425 1372]{ATLAS_SUSY_Strong_all.png}
%\hspace{2.0cm}
\vspace{-0.6cm}
\caption{ Summary of the exclusion limit on the gluino decay\cite{summaryplot_gluino}. }
\label{fig:summary_limit}
\end{figure}

The ATLAS collaboration has performed many searches for gluinos and squarks using the 2015+2016 dataset collected at $\sqrt{s}=$13~TeV. 
None of the searches have shown a significant deviation from the Standard Model predictions.
Figure \ref{fig:summary_limit} shows a summary of the exclusion limits in the mass plane of gluino and LSP for different searches featuring the decay of the gluino to the LSP. 

However, the experiment of $\sqrt{s}=$13~TeV collision has just started. We will have a large increase on the sensitivity in coming data.

%\vspace{2cm}

%%%%%%%%%%%%%%%%%%%%%%%%%%%%%%%%%%%%%%%%%%%%%%%%%%%%%%%%%%%%%%%%%%%%%%%%%
%%
%%   use this format to include an .eps figure into your paper
%%
%\begin{figure}[htb]
%\centering
%\includegraphics[width=0.9\textwidth]{head_lhcp2017.pdf}
%\includegraphics[height=2in,bb=0 0 713 251]{head_lhcp2017.jpg}
%\caption{ Place the caption here}
%\label{fig:figure1}
%\end{figure}
%%%%%%%%%%%%%%%%%%%%%%%%%%%%%%%%%%%%%%%%%%%%%%%%%%%%%%%%%%%%%%%%%%%%%%%%%%%

%See Figure \ref{fig:figure1} and Table \ref{tab:table1}. 

%%%%%%%%%%%%%%%%%%%%%%%%%%%%%%%%%%%%%%%%%%%%%%%%%%%%%%%%%%%%%%%%%%%%%%%%%
%%
%%   use this format to include a LaTeX table  into your paper
%%
%\begin{table}[t]
%\begin{center}
%\begin{tabular}{l|ccc}  
%Patient &  Initial level($\mu$g/cc) &  w. Magnet &  
%w. Magnet and Sound \\ \hline
% Guglielmo B.  &   0.12     &     0.10      &     0.001  \\
% Ferrando di N. &  0.15     &     0.11      &  $< 0.0005$ \\ \hline
%\end{tabular}
%\caption{ place the caption here }
%\label{tab:table1}
%\end{center}
%\end{table}
%%%%%%%%%%%%%%%%%%%%%%%%%%%%%%%%%%%%%%%%%%%%%%%%%%%%%%%%%%%%%%%%%%%%%%%%%%%

%%  if necessary
%\Acknowledgements
%I am grateful to XYZ for fruitful discussions.

\end{document}